\documentclass[sigconf]{acmart}
\renewcommand\footnotetextcopyrightpermission[1]{} 

\setcopyright{none}
\AtBeginDocument{%
  }

\usepackage{graphicx}
\usepackage{subcaption}

\begin{document}

\title{What Can Eye Gaze Teach Us About Real-World Cycling? Insights From the Oxford RobotCycle Project}

\author{Benjamin Hardin}
\orcid{0009-0000-9938-3336}
\affiliation{%
  \institution{University of Oxford}
  \city{Oxford}
  \country{United Kingdom}
}
\email{benjamin.hardin@cs.ox.ac.uk}

\author{Efimia Panagiotaki}
\orcid{0009-0008-0189-8904}
\affiliation{%
  \institution{University of Oxford}
  \city{Oxford}
  \country{United Kingdom}
}
\email{efimia@robots.ox.ac.uk}

\author{Daniele De Martini}
\orcid{0000-0001-6121-5839}
\affiliation{%
  \institution{University of Oxford}
  \city{Oxford}
  \country{United Kingdom}
}
\email{daniele@robots.ox.ac.uk}

\author{Lars Kunze}
\orcid{0000-0001-5302-1938}
\affiliation{%
  \institution{University of the West of England Bristol}
  \city{Bristol}
  \country{United Kingdom}
}
\email{lars.kunze@uwe.ac.uk}

\renewcommand{\shortauthors}{Hardin et al.}

\begin{abstract}
Although much is known about the physical danger of cycling situations, less is understood about the perceived danger of cycling. Furthermore, perception of danger may be filtered at a subconscious level and therefore difficult for one to self-report. To this end, these subconscious perceptions can be revealed through physiological metrics such as eye gaze. 
This paper explores the perceived safety of cycling in Oxford, United Kingdom and explores the ability of wearable eye tracking glasses to produce insights about the differences in perception under different environments and events.
This paper finds that eye gaze patterns change between using bike lanes, car lanes and shared bus lanes, representing different cognitive challenges of each lane type.
This paper presents that different intersections have significantly different eye gaze patterns which may have implications for cyclist stress. Finally, eye gaze patterns differ in the presence of events such as passes and pedestrians in the road compared to when cycling with no events.
This paper draws conclusions on the benefits and limitations of using wearable eye trackers to estimate stress and cyclist workload.
\end{abstract}




\maketitle

\pagestyle{plain} 

\section{Introduction}
Cycling continues to be a popular form of transportation that has seen significant interest in recent years due to health and environmental benefits as well as enjoyment.
In order to ensure that cycling is a safe form of transportation, it is important to understand not only situations where accidents occur, but also real time responses of cyclists to various events and environment factors. Nevertheless, it is difficult to collect real-time in-situ feedback from cyclists because they must focus full attention on the task. Many prior studies have had to rely on post-hoc or simulated environment responses from cyclists which means many situations and smaller interactions may be forgotten \cite{GADSBY2021102932, RIVERAOLSSON2023107007}. To this end, there is potential for measuring physiological metrics such as heart rate, heart rate variability, and skin response in real time and using these metrics as a proxy for stress and workload \cite{10457965, CAVIEDES2018488}. Although perhaps less explored, tracking eye gaze is another potential physiological metric that can be used to measure stress and excitement and has already shown some application to cycling studies \cite{MA202452}.

In this paper, we present findings from the RobotCycle trials in Oxford, UK conducted over several days in 2024 and 2025. We measure eye gaze fixations and horizontal gaze dispersion to understand scenarios, routes, and events that elicit the greatest reaction from the cyclist and explore the feasibility of eye gaze for measuring cyclist stress. We find that eye gaze metrics differ significantly between various intersection types, that various events that the cyclist may encounter (such as overtakes or pedestrians) produce significantly different eye gaze patterns compared to no events, and that different routes demonstrate significantly different eye gaze patterns due to varying infrastructures, traffic, and pedestrian presence. However, due to the complexity of environments and their multitude of elements which may capture visual attention, we highlight the limitations that eye gaze alone has on understanding an environment. We find that eye gaze patterns alone were not sufficient to distinguish between types of events that a cyclist faces. For these situations, object detection and segmentation is necessary. Furthermore, fixation duration and horizontal gaze dispersion alone may be insufficient for reliably estimating cyclist stress.

\section{Literature Review}

\subsection{Relationship Between Eye Gaze and Stress}
Eye gaze movements have been used extensively to evaluate stress, fatigue, and emotions \cite{yousefi2022, li2021, bend2026comparing, yang2025} across diverse domains including technical interviews, student learning, driving, and nursing \cite{behroozi2018can, Jyotsna, nemcova, ahmadi2024quantifying}. There have also been numerous methods proposed for measuring eye gaze metrics, with the most popular ones including velocity threshold (I-VT) and dispersion threshold (I-DT) methods \cite{andersson2017one, salvucci2000identifying}. Nevertheless, reliability of using eye gaze to predict psychological factors has been shown to be very dependent on the context and scenario in which the eye gaze is being measured. Intuitively, some tasks may require more scanning across a broad environment while other tasks involve deep focus on a particular point or set of points. For instance, one paper considers solving technical interview problems on a whiteboard compared to on paper and uses this as a basis of demonstration that gaze metrics vary in their meanings \cite{behroozi2018can}. Additionally, there is a question about the reliability of eye gaze in non-static environments such as walking, skipping, or jumping \cite{hooge2023robust, franchak2021adapting}. To this end, we compare eye gaze across environment factors, establishing the eye gaze baseline as the presence of no events or manoeuvres. 

\subsection{Risky Situations in Cycling}
One study has found that situations that are perceived as risky generally align with the situations that are objectively risky \cite{von2020crash}. Furthermore, the study found that bike lanes did not reliably reduce crash risks and that public transport stops could increase risk. Higher traffic volume and greater intersection complexity were also related to increased crash risks, a finding that was corroborated by a study of cycling in London \cite{aldred2018cycling}.

It has been found that intersections can induce higher levels of stress for cyclists \cite{CAVIEDES2018488}. The authors also found that protected bike lanes were associated with reduced stress levels. One study also found that discontinuities in paths, such as cycle paths ending, could increase stress as well as situations when there is not enough space for bicycles \cite{su10072379}. High speed differentials between cyclists and vehicles have also been found to be stressful \cite{GADSBY2021102932}. 

\section{Eye Gaze and Cycling}
The systematic review by Ma et al., 2024 presents an overview of eye gaze tracking in cycling studies \cite{MA202452}, finding that there is a lack of standardised practice for variables such as glances and fixations and highlights that fixations may not consider all of the data that cyclists use from their peripheral vision. Furthermore, the paper considers how types of workload may vary extensively by environment but this has yet to be studied.
One study considered eye gaze patterns of cyclists at intersections \cite{rupi2019}, finding that intersections which force cyclists to share a lane with vehicles induce more eye gaze patterns associated with stress. Another study found that the presence of pedestrians significantly disrupts a cyclist's fixation on the path before them as well as when there are changes to the cycling path \cite{MANTUANO2017408}. The paper proposes the need for greater visual separation between pedestrians and cyclists when they share existing infrastructure. One real-world study of cyclists found that subjective risk was related to the spatial complexity of the environment, and this complexity contributed to more hectic gaze behaviour, shorter fixation durations, and more fixations away from the direction of travel \cite{von2020gaze}. Most importantly, the study found that how far ahead cyclists looked and how far their gaze deviated from the direction of travel are more informative measures of perceived danger than fixation durations. Finally, one paper has found that right turns for cyclists (in a right-hand traffic context) can induce greater mental workload than left turns \cite{rawson2025eyes}.

\def\rqone{How do eye gaze metrics compare when the cyclist is utilising a bike, bus, or car lane?}
\def\rqtwo{How does eye gaze compare for roundabouts, signalised crossings, unsignalised turns, and zebra crossings?}
\def\rqthree{How do eye gaze metrics compare across events (as defined in Table~\ref{tab:annotation_labels})?}
\def\rqfour{How do cyclists respond to passes when a bike lane is present versus when no bike lane is present?}
\def\rqfive{How does cyclist eye gaze compare across routes?}

\section{Methodology}
\subsection{Research Questions}
The research is motivated by the following questions:
\begin{enumerate}
    \item[\textbf{RQ1.}] \rqone
    \item[\textbf{RQ2.}] \rqtwo
    \item[\textbf{RQ3.}] \rqthree
    \item[\textbf{RQ4.}] \rqfive
\end{enumerate}

\subsection{Data Collection}
\begin{figure}[h]
    \centering
    \begin{subfigure}{0.20\textwidth}
        \includegraphics[trim=490 75 450 150, clip,width=\textwidth]{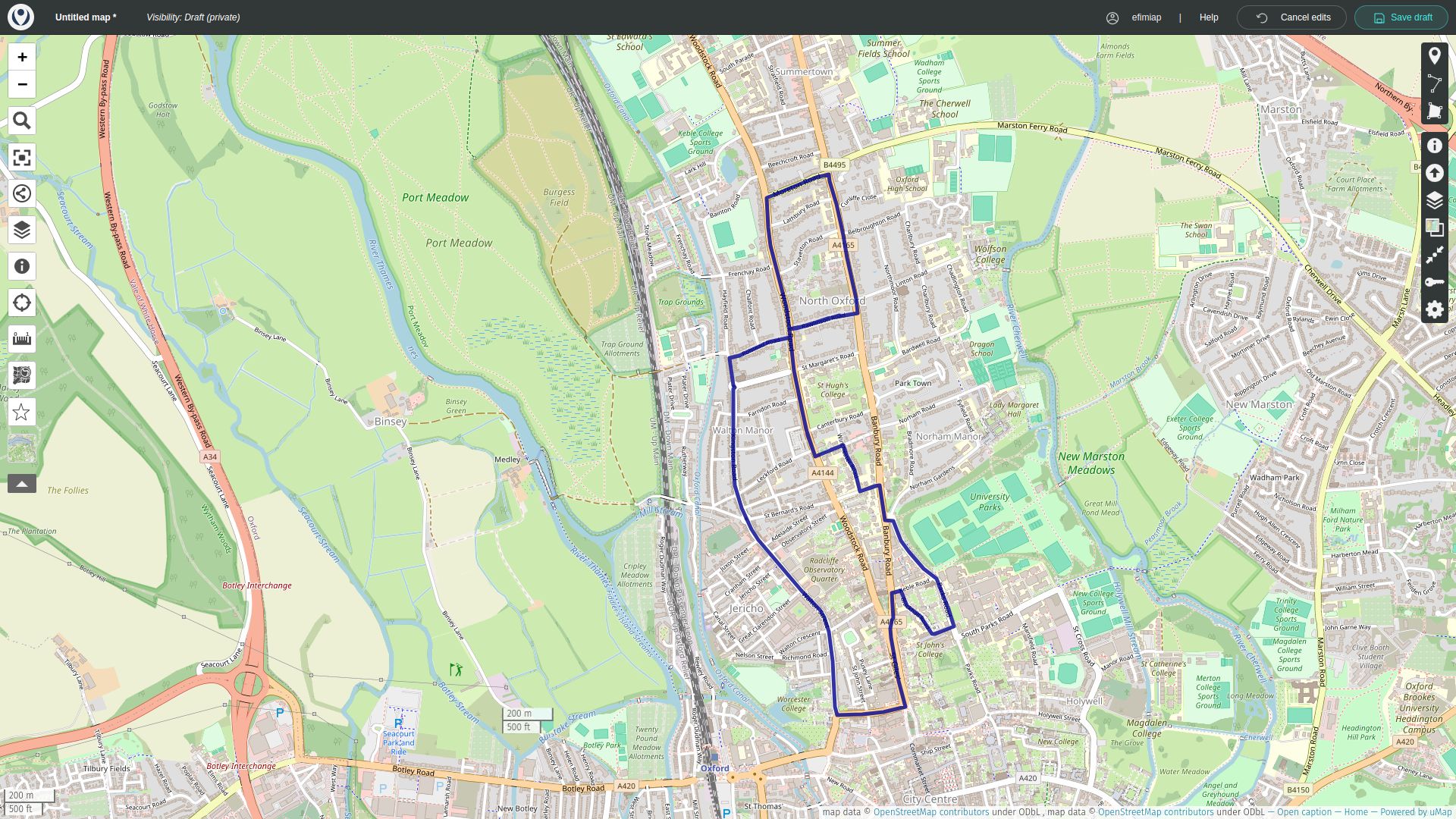}
        \caption{North Loop: 6.05 km}
    \end{subfigure}
    \begin{subfigure}{0.20\textwidth}
        \includegraphics[trim=490 75 450 150, clip,width=\textwidth]{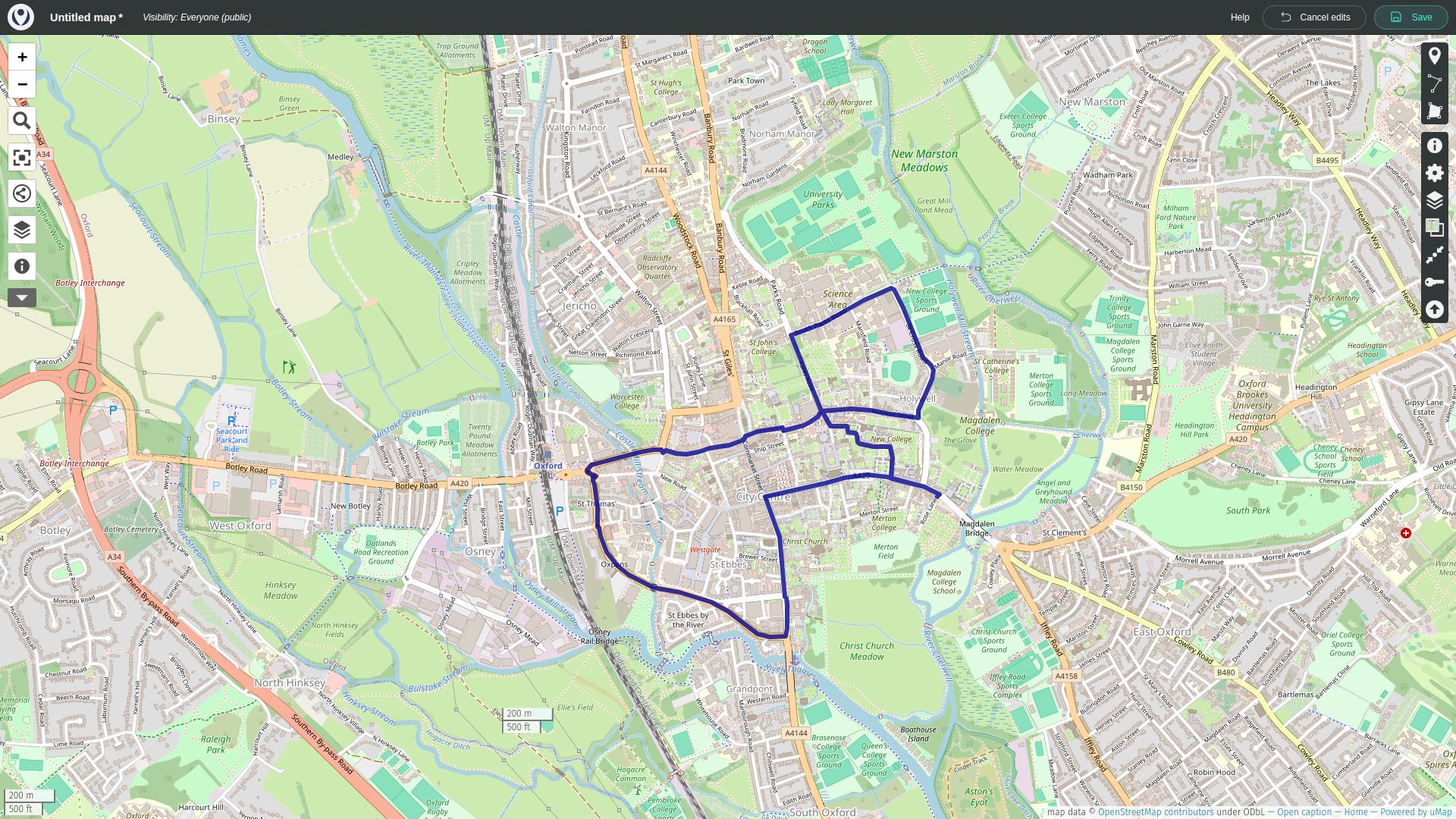}
        \caption{Central Loop: 5.38 km}
    \end{subfigure}
    \begin{subfigure}{0.20\textwidth}
        \includegraphics[trim=490 75 450 150, clip,width=\textwidth]{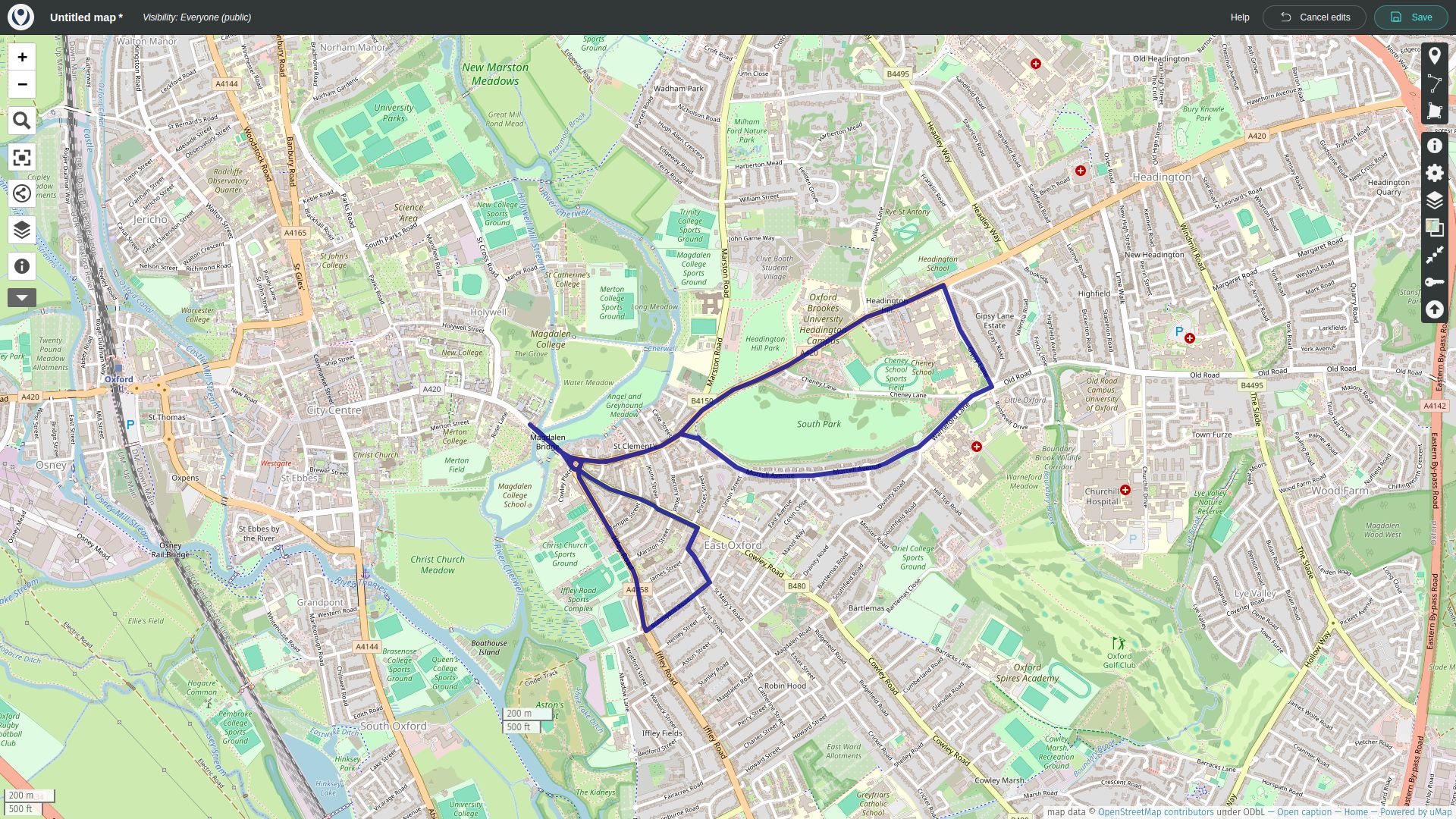}
        \caption{South Loop: 5.99 km}
    \end{subfigure}
    \caption{Data Collection Routes in Oxford.}
    \label{fig:route_plan}
\end{figure}

Data was collected using the first-generation Meta Aria research glasses \cite{aria} which include two eye tracking cameras, two external-facing mono scene cameras, and one POV RGB camera. A cyclist wore the glasses while cycling around 3 different routes in Oxford over the course of several days in 2024 and 2025. An overview of the routes is presented in Figure~\ref{fig:route_plan}. 
Routes were cycled in both directions and routes demonstrate low-speed environments of 20 and 30 mph roads through city centre and urban neighbourhood areas. Data was collected from 12 runs (5 city centre, 4 north, 3 south). Each route generally took 20-30 minutes to complete, allowing ample time for diverse situations to be encountered including emergency vehicles, close vehicle passes, and near misses with jaywalking pedestrians. The cyclist also wore the RobotCycle backpack \cite{robotcycle} which contains a number of sensors for monitoring the environment. While data from this backpack is not included in this paper, it is worth noting as this backpack may influence cyclist agility, speed, and manoeuvring.
Data was collected and handled in accordance with the University of Oxford central privacy and ethics approval.

\subsection{Analysis}
Eye gaze projection points onto the scene were calculated using the Meta Aria Research Toolkit \cite{aria}.
To estimate eye movement patterns, a variation of the popular I-VT algorithm \cite{salvucci2000identifying} was utilised using the pymovements python package version 0.26.2 \cite{pymovements}.
Segments where the bike was stationary at traffic lights, before they started the route, or after they reached the finish were excluded as the cyclist is not focused on the riding during these moments.

To analyse fixation durations and horizontal gaze dispersions across the environment variables in our research questions, we first log-transformed our data to address a strong-right skew present in the distributions. We then performed Welch's ANOVA on the log-transformed data as an omnibus test followed by Games-Howell post-hoc comparisons for more detailed analysis between variables. In our results, the \textit{p} value represents the value from the Games-Howell post-hoc comparisons, the \textit{g} value represents Hedges' g, and the \textit{ratio} value represents the geometric mean ratio of the variables.

\subsection{Data Annotation}
A summary of annotations labels can be found in Table~\ref{tab:annotation_labels}. Labels that should be noted include \textit{ego\_passing} which involves the cyclist overtaking a vehicle that was parked in a position that is not a normal parking spot and required the cyclist to adjust their trajectory for it. A \textit{pass} occurs whenever a vehicle passes the cyclist in the same direction as the cyclist is travelling and a \textit{close\_pass} was labelled when the vehicle passed closer than would be expected given the available road space. There were few close passes in the final data. The \textit{oncoming\_pass} label represents a vehicle travelling in the opposite direction of the cyclist and was only an annotated event when the road was narrow or the lane was shared for both directions. Thus, an oncoming pass from a vehicle in an opposite lane separated by a lane line was not annotated. Passes and oncoming pass events only apply to vehicles, motorcycles, and scooters and do not include pedestrians or other cyclists. Passes were annotated starting at the moment they appear in the video, however, the cyclist is likely to have heard the vehicle passing from behind. Thus, we perform analysis on a few seconds preceding the pass. A \textit{vehicle\_pull\_out\_front} annotation was applied whenever a vehicle crossed the cyclist's path in the near distance in a way that caused the cyclist to reduce speed or change trajectory. Data was annotated using ELAN 7.1 \cite{elan}. Annotation was performed by one researcher to ensure maximum consistency in annotation style across samples.

\begin{table}[]
    \centering
    \caption{Route Annotation Labels}
    \begin{tabular}{l l}
    \hline
         \textit{infrastructure\_type} & bike\_lane\\
          & car\_lane \\
          & shared\_bus\_lane \\
          & bike\_path \\
          & shared\_path \\
          & off\_road \\
          \hline
         \textit{event} & ego\_passing \\
         & pass \\
         & close\_pass \\
         & oncoming\_pass \\
         & pedestrian\_in\_lane \\
         & vehicle\_pull\_out\_front \\
         \hline
         \textit{intersection\_context} & roundabout \\
         & passing\_signalised\_crossing \\
         & left\_turn \\
         & right\_turn \\
         & zebra\_crossing \\
         & crossing\_road \\
         & lane\_change \\
         \hline
         \textit{other\_context} & bike\_stopped \\
         & construction\_zone \\
         & object\_in\_road \\
         & emergency\_vehicle 
    \end{tabular}
    \label{tab:annotation_labels}
\end{table}

For \textit{infrastructure\_type}, the data is labelled based on where the bike is in the scene, not what infrastructure is available. For instance, a given segment may have a bike lane available but the cyclist was in a turn lane instead. For this segment, the data would be labelled as \textit{car\_lane}. A \textit{bike\_path} represents a lane only available to bikes and with a median between other lanes such as pedestrian paths or car lanes. A \textit{shared\_path} represents a path that is used by both pedestrians and cyclists.

\section{Results}
\subsubsection{RQ1. \rqone}
Throughout Oxford and particularly on the chosen routes, bike lanes are mostly very narrow (the width of a single bike), unprotected, and separated from car lanes by broken lines rather than solid lines. Because of this, we expect that the presence of a bike lane does not have a meaningful impact on the perceived safety compared to a car lane. Bus lanes, however, are quite broad and provide a wide space between the cyclist and other vehicles, meaning we should expect a measurable difference in eye gaze metrics.

\begin{figure}[h]
  \centering
  \caption{Fixation Duration by Infrastructure Type}
  \includegraphics[width=0.8\linewidth]{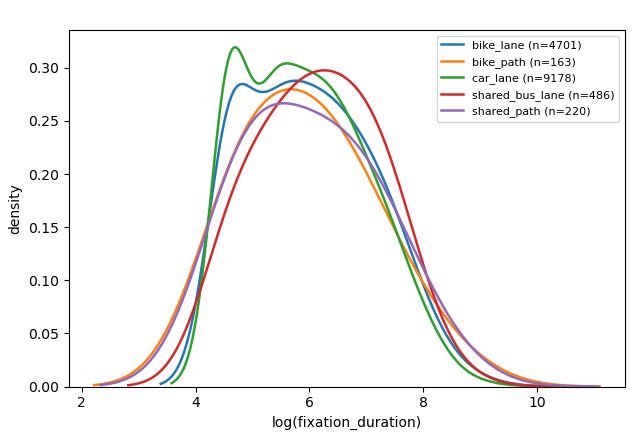}
  \label{fig:fixation_duration_infrastructure_type}
\end{figure}

The strongest effect of fixation duration occurs between car and bus lanes (p<0.001, g=-0.21, ratio=0.80), where the fixation durations are about 20\% shorter in the car lane. Another significant effect occurs between bike and car lanes (p=0.001, g=0.07, ratio=1.08), where fixations are shorter during bike lane riding. However, the difference is only 8\% and the sample size is large, which may mean the significant effect is not as strong as it appears.
When comparing bike lanes to bus lanes, there is a significant effect (p=0.023, g=-0.14, ratio=0.86), representing about 14\% shorter fixation durations in bike lanes than bus lanes. Density of fixation durations by infrastructure type can be found in Figure~\ref{fig:fixation_duration_infrastructure_type}.

\begin{figure}[h]
  \centering
  \caption{Dispersion by Infrastructure Type}
  \includegraphics[width=0.8\linewidth]{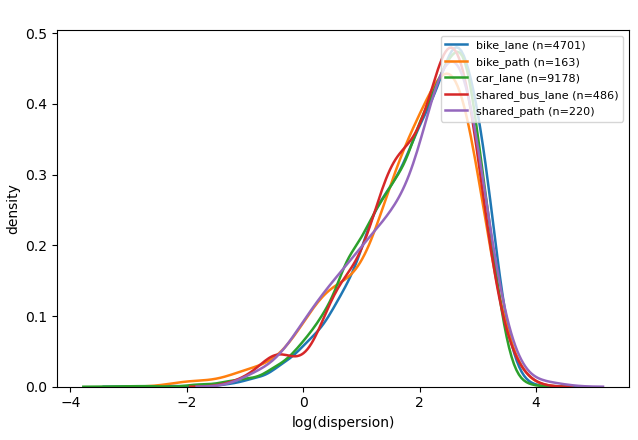}
  \label{fig:dispersion_infrastructure_type}
\end{figure}

Concerning the strongest effect of infrastructure on dispersion, dispersion was approximately 8.5\% larger during bike lane riding compared to car lane riding (p<0.001, g=0.09, ratio=1.085), a similar mean difference to what was found with fixations for these two lane types. No other significant effects were found between infrastructure types on dispersion. Density of dispersions by infrastructure type can be found in Figure~\ref{fig:dispersion_infrastructure_type}.

These findings are not surprising given the narrow nature of most Oxford bike lanes, leaving the cyclist almost as exposed to cars as if they were in the car lane. Thus, the similarities between bike and car lanes that we see in fixation and dispersion comparison align with expectations. With the large width and general solid painted stripe of the bus lane, along with bus drivers in Oxford who are very used to navigating around cyclists, this leaves more time for the cyclist to fixate on the environment and requires less scanning behaviour as is seen in the greater fixation durations of the bus lane.

We also calculated fixation frequency, measured in fixation events per minute, and found that bike lanes had the highest frequency at 142.2 fixations per minute, followed by car lanes at 80.5 fixations per minute, and finally bus lanes at 62.8 fixations per minute. These results combined with the dispersion results indicate that when in bike lanes, the cyclist is able to more rapidly move their fixation on points in the environment, possibly representing increased cognitive load as they rapidly switch between fixation points. Since the bus lane demonstrated the longest fixation duration times, it is not surprising that it also demonstrates the lowest fixation frequency. 

\subsubsection{RQ2. \rqtwo}
Although labelled in the dataset, we exclude analysis during scenarios where the cyclists crossed another road via a yield because this intersection type occurred very infrequently, only once per run in the south loop.

\begin{figure}[h]
  \centering
  \caption{Fixation Duration by Intersection Type}
  \includegraphics[width=0.8\linewidth]{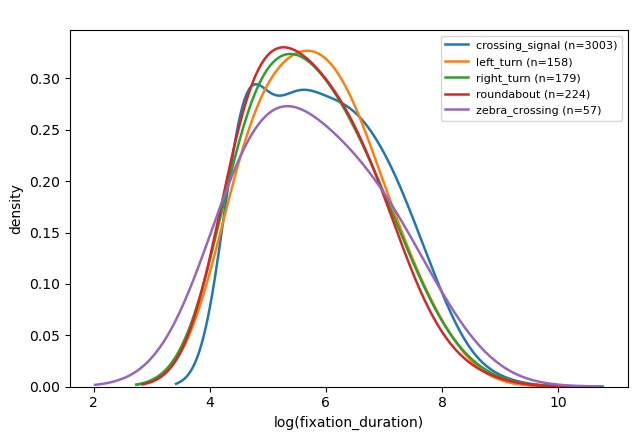}
  \label{fig:fixation_duration_intersection}
\end{figure}

Passing a signalised crossing shows significance compared to roundabouts (p=0.004, ratio=1.27, g=0.22), representing that fixation durations are about 27\% longer during signalised crossings compared to roundabouts. At near significance (p=0.065), fixations at signalised crossings are 22\% longer than during right turns as well (ratio=1.22, g=0.19). Although not significant, signalised crossings produce longer fixations overall than every other intersection type. The density distribution of fixation durations by intersection type is presented in Figure~\ref{fig:fixation_duration_intersection}.

This may indicate that when passing signalised crossings, fixations occur on specific elements (such as the traffic signal itself or pedestrians waiting to cross), while other intersections require rapid scanning of the environment for vehicles entering and exiting as well as the desired exit point, leading to shorter individual fixations as gaze shifts more frequently. This is likely particular to Oxford where many of the signalised crossings are only for pedestrians and do not contain joining or crossing roads at the signal. In those scenarios, there is less of a burden for the cyclist to scan for traffic that may join from other directions.

\begin{figure}[h]
  \centering
  \caption{Dispersion by Intersection Type}
  \includegraphics[width=0.8\linewidth]{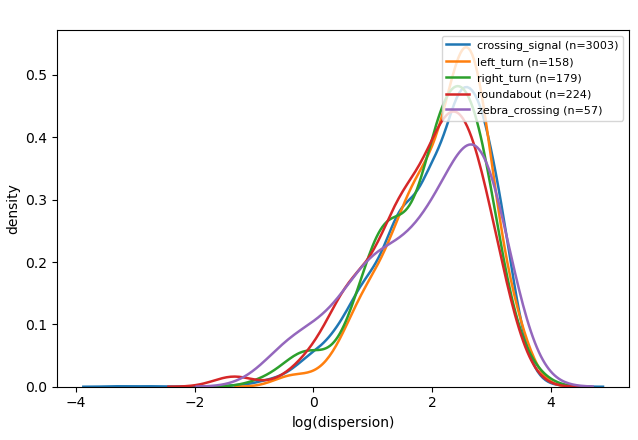}
  \label{fig:dispersion_intersection}
\end{figure}

Concerning gaze dispersion, we find a significant effect for left turns compared to roundabouts (p=0.031, ratio=1.29, g=0.29), where gaze dispersion during left turns is 29\% greater than during roundabouts. Signalised crossings compared to roundabouts represent a near significant effect (p=0.057, ratio=1.19, g=0.19), where gaze dispersion during signalised crossings is 19\% greater than during roundabouts. No other pairwise comparisons reached or were near significance. This may indicate that roundabouts allow for more fixed gaze as there is more constrained flow of traffic compared to left turns or places where pedestrians are likely to cross. Overall, roundabouts elicit a noticeably different gaze pattern compared to other intersection types, as roundabouts represent the shortest fixations and the most focused gaze. The density distribution by intersection type is presented in Figure~\ref{fig:dispersion_intersection}.

\subsubsection{RQ3. \rqthree}

In general, there was no statistically significant effect in fixation duration between event types (Welch's ANOVA p=0.586). The Games-Howell post-hoc comparison confirmed this finding, as no pairwise comparison achieved significance (p<0.05). Geometric mean fixation durations across the event types are very consistent: roughly 354 ms (\textit{pedestrian\_in\_lane}), 346 ms (\textit{vehicle\_pull\_out\_front}), 343 ms (\textit{oncoming\_pass}), 357 ms (\textit{ego\_passing}), 379 ms (\textit{pass}). Across thousands of fixations and five event types, the spread is only about 35 ms.

\begin{figure}[h]
  \centering
  \caption{Fixation Duration by Event}
  \includegraphics[width=0.8\linewidth]{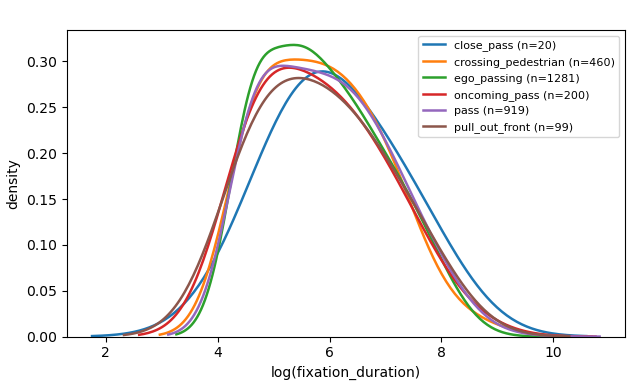}
  \label{fig:fixation_event}
\end{figure}

\begin{figure}[h]
  \centering
  \caption{Dispersion by Event}
  \includegraphics[width=0.8\linewidth]{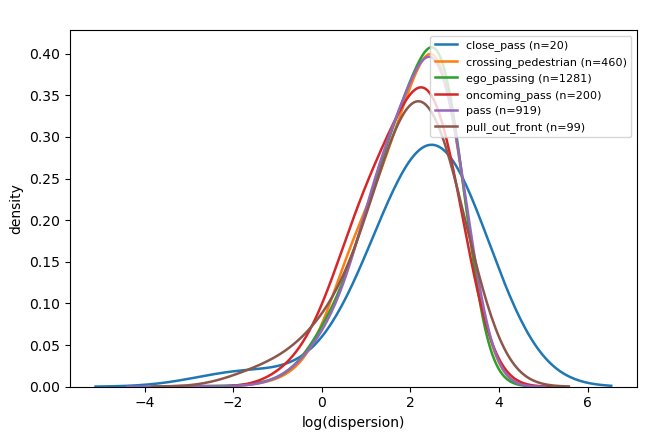}
  \label{fig:dispersion_event}
\end{figure}

We performed the analysis for gaze dispersion metrics across events and found a similar null result (Welch's ANOVA p=0.207). The Games-Howell post-hoc comparison confirmed this finding, as no pairwise comparison achieved significance (p<0.05). Nevertheless, there is a 23\% difference in dispersion across events. Events involving lateral spatial relationships (\textit{close\_pass, pass, pedestrian\_in\_lane, ego\_passing}) tend toward more dispersed gaze, while events ahead in the road (\textit{oncoming\_pass, vehicle\_pull\_out\_front}) tend toward more focused gaze. Lateral events require scanning across the cyclist's flanks (checking adjacent vehicles, monitoring overtaking, tracking pedestrians on cross-paths), which intuitively would result in broader gaze dispersions. Events which require attention to a single point ahead, such as the oncoming vehicle or the vehicle pulling out, should produce more focused gaze. Nevertheless, this pattern does not reach significance. Densities of fixation durations by event and dispersions by event are presented in Figure~\ref{fig:fixation_event} and Figure~\ref{fig:dispersion_event}, respectively.

When we compared fixations during event periods (n = 2979) to non-event periods, we found that a Welch's ANOVA on log-transformed durations revealed a statistically significant difference (p<0.001); however, the effect was extremely small in magnitude (g=0.08).
Geometric mean fixation duration was approximately 393 ms during non-event periods compared to 362 ms during events, a difference of roughly 31 ms, or 8.7\% (95\% CI: 4.2\% to 13.4\%). This finding indicates that gaze patterns do indeed change in the presence of events, but high variation in eye gaze patterns means we cannot distinguish between types of events from gaze metrics alone.



\subsubsection{RQ4. \rqfive}
\begin{figure}[h]
  \centering
  \caption{Fixation Duration by Route}
  \includegraphics[width=0.8\linewidth]{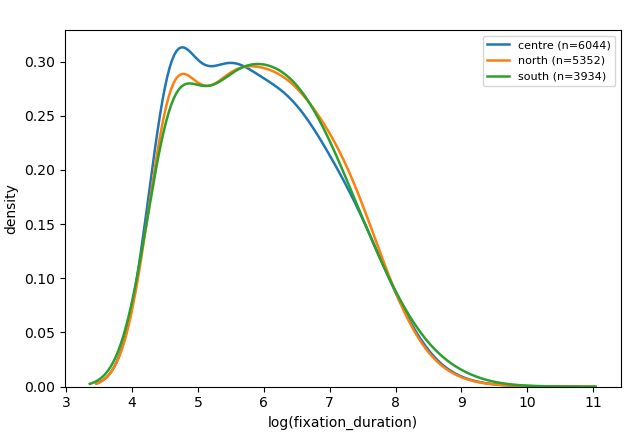}
  \label{fig:fixation_by_route}
\end{figure}

\begin{figure}[h]
  \centering
  \caption{Gaze Dispersion by Route}
  \includegraphics[width=0.8\linewidth]{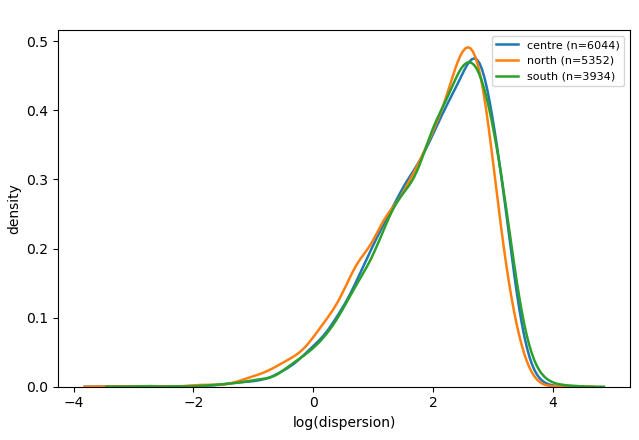}
  \label{fig:dispersion_by_route}
\end{figure}

The south route had the highest average fixation duration at 754 ms, followed by north at 710 ms, and the centre route at 692 ms. These results intuitively make sense as the southern route contains much longer straight stretches of road. Figure~\ref{fig:fixation_by_route} and Figure~\ref{fig:dispersion_by_route} detail the density distribution of fixations and dispersions, respectively, by routes. 

There is a statistically significant effect between fixation durations for the centre and north routes (p=0.019, g=-0.05, ratio=0.95) and between the centre and south routes (p=0.005, g=-0.06, ratio=0.93). However, there is no significant effect between the north and south routes (p=0.767, g=-0.015, ratio=0.98).

Concerning gaze dispersion, there is a significant effect between the centre and north route (p<0.001, g=0.11, ratio=1.11) and between the north and south route (p<0.001, g=-0.14, ratio=0.88). There is no significant effect between the centre and south route (p=0.357, g=-0.03, ratio=0.97). Overall, the north route elicits more focused gaze than both the centre and south routes.



\section{Discussion and Conclusion}
This study measured the eye gaze of a cyclist across diverse routes in Oxford, UK under different traffic and weather conditions across 2024 and 2025. The routes represent a broad mix of bike lanes, shared bike lanes, bike paths, bus lanes, bus traffic, car traffic, and pedestrian presence.

Our results indicate that raw fixation durations alone proved not to be a useful discriminator across event types and thus fixation duration alone is not enough to predict an event type. These results highlight the importance of understanding which objects the cyclist fixates on at which points in time. This will be an important area of future research to improve the usefulness of the dataset. Nevertheless, fixation durations and gaze dispersions provide useful insights about cyclist responses to infrastructure, notably that roundabouts generally allow for more focused attention due to the constrained flow of traffic.

There are a few limitations to the present study. Future work may consider the inclusion of inertial measurement unit (IMU) data in the analysis to improve the understanding of fixations in relation to head movements. Furthermore, future work should use the gaze projection points on the environment to understand not just what is in the scene but what the cyclist is directly looking at during fixations. Understanding not only when the cyclist fixates, but what the cyclist fixates on could further enhance the insights about what triggers a fixation in diverse conditions and environments.

Our study reveals the importance of the metrics of fixation duration and horizontal gaze dispersion for understanding how cyclists perceive and interact with various traffic environments. However, our results highlight the limited ability of these two metrics alone to predict stress or workload of a rider. 

\begin{acks}
The authors would like to acknowledge the immense work of the RobotCycle and Oxford Robotics Institute team, particularly Pratik Somaiya and Daniel Marques, for organising, collecting, and anonymising the data that was used for this project.
\end{acks}

\bibliographystyle{ACM-Reference-Format}
\bibliography{references}

@ARTICLE{robotcycle,
  author={Panagiotaki, Efimia and Thuremella, Divya and Baghabrah, Jumana and Sze, Samuel and Frank Tarimo Fu, Lanke and Hardin, Benjamin and Reinmund, Tyler and Flatscher, Tobit and Marques, Daniel and Prahacs, Chris and Kunze, Lars and de Martini, Daniele},
  journal={IEEE Transactions on Field Robotics}, 
  title={The Oxford RobotCycle Project: A Multimodal Urban Cycling Dataset for Assessing the Safety of Vulnerable Road Users}, 
  year={2025},
  volume={2},
  number={},
  pages={308-335},
  doi={10.1109/TFR.2025.3566304}}

@misc{elan,
  author       = {{Nijmegen: Max Planck Institute for Psycholinguistics}},
  title        = {{ELAN (Version 7.1)}},
  year         = {2026},
  howpublished = {Computer software},
  url          = {https://archive.mpi.nl/tla/elan},
  note         = {Retrieved from https://archive.mpi.nl/tla/elan}
}

@article{aria,
  title={Project aria: A new tool for egocentric multi-modal ai research},
  author={Engel, Jakob and Somasundaram, Kiran and Goesele, Michael and Sun, Albert and Gamino, Alexander and Turner, Andrew and Talattof, Arjang and Yuan, Arnie and Souti, Bilal and Meredith, Brighid and others},
  journal={arXiv preprint arXiv:2308.13561},
  year={2023}
}

@inproceedings{pymovements,
    author = {Krakowczyk, Daniel G. and Reich, David R. and Chwastek, Jakob and Jakobi, Deborah N.
 and Prasse, Paul and Süss, Assunta and Turuta, Oleksii and Kasprowski, Paweł
 and Jäger, Lena A.},
    title = {pymovements: A Python Package for Processing Eye Movement Data},
    year = {2023},
    isbn = {979-8-4007-0150-4/23/05},
    publisher = {Association for Computing Machinery},
    address = {New York, NY, USA},
    url = {https://doi.org/10.1145/3588015.3590134},
    doi = {10.1145/3588015.3590134},
    booktitle = {2023 Symposium on Eye Tracking Research and Applications},
    location = {Tubingen, Germany},
    series = {ETRA '23},
}

@article{CAVIEDES2018488,
title = {Modeling the impact of traffic conditions and bicycle facilities on cyclists’ on-road stress levels},
journal = {Transportation Research Part F: Traffic Psychology and Behaviour},
volume = {58},
pages = {488-499},
year = {2018},
issn = {1369-8478},
doi = {https://doi.org/10.1016/j.trf.2018.06.032},
url = {https://www.sciencedirect.com/science/article/pii/S1369847817305466},
author = {Alvaro Caviedes and Miguel Figliozzi}
}

@Article{su10072379,
AUTHOR = {Nuñez, Javier Yesid Mahecha and Teixeira, Inaian Pignatti and Silva, Antônio Nélson Rodrigues da and Zeile, Peter and Dekoninck, Luc and Botteldooren, Dick},
TITLE = {The Influence of Noise, Vibration, Cycle Paths, and Period of Day on Stress Experienced by Cyclists},
JOURNAL = {Sustainability},
VOLUME = {10},
YEAR = {2018},
NUMBER = {7},
ARTICLE-NUMBER = {2379},
URL = {https://www.mdpi.com/2071-1050/10/7/2379},
ISSN = {2071-1050},
DOI = {10.3390/su10072379}
}

@Article{rupi2019,
AUTHOR = {Rupi, Federico and Krizek, Kevin J.},
TITLE = {Visual Eye Gaze While Cycling: Analyzing Eye Tracking at Signalized Intersections in Urban Conditions},
JOURNAL = {Sustainability},
VOLUME = {11},
YEAR = {2019},
NUMBER = {21},
ARTICLE-NUMBER = {6089},
URL = {https://www.mdpi.com/2071-1050/11/21/6089},
ISSN = {2071-1050},
DOI = {10.3390/su11216089}
}

@article{MANTUANO2017408,
title = {Cyclist gaze behavior in urban space: An eye-tracking experiment on the bicycle network of Bologna},
journal = {Case Studies on Transport Policy},
volume = {5},
number = {2},
pages = {408-416},
year = {2017},
issn = {2213-624X},
doi = {https://doi.org/10.1016/j.cstp.2016.06.001},
url = {https://www.sciencedirect.com/science/article/pii/S2213624X16300220},
author = {Alessandra Mantuano and Silvia Bernardi and Federico Rupi}
}

@article{MA202452,
title = {Eye tracking measures of bicyclists’ behavior and perception: A systematic review},
journal = {Transportation Research Part F: Traffic Psychology and Behaviour},
volume = {107},
pages = {52-68},
year = {2024},
issn = {1369-8478},
doi = {https://doi.org/10.1016/j.trf.2024.08.026},
url = {https://www.sciencedirect.com/science/article/pii/S136984782400233X},
author = {Shiyu Ma and Wenwen Zhang and Robert B. Noland and Clinton J. Andrews}
}

@article{rawson2025eyes,
  title={Eyes on the Road: Disentangling Cyclist Mental Workload at Intersections},
  author={Rawson, James Joseph},
  year={2025}
}

@ARTICLE{10457965,
  author={Guo, Xiang and Tavakoli, Arash and Chen, T. Donna and Heydarian, Arsalan},
  journal={IEEE Transactions on Intelligent Transportation Systems}, 
  title={Unveiling the Impact of Cognitive Distraction on Cyclists Psycho-Behavioral Responses in an Immersive Virtual Environment}, 
  year={2024},
  volume={25},
  number={8},
  pages={10274-10285},
  doi={10.1109/TITS.2024.3366777}}

@ARTICLE{yousefi2022,
  author={Yousefi, Mansoureh Seyed and Reisi, Farnoush and Daliri, Mohammad Reza and Shalchyan, Vahid},
  journal={IEEE Access}, 
  title={Stress Detection Using Eye Tracking Data: An Evaluation of Full Parameters}, 
  year={2022},
  volume={10},
  number={},
  pages={118941-118952},
  doi={10.1109/ACCESS.2022.3221179}}

@INPROCEEDINGS{li2021,
  author={Li, Fan and Xu, Gangyan and Feng, Shanshan},
  booktitle={2021 IEEE International Conference on Systems, Man, and Cybernetics (SMC)}, 
  title={Eye Tracking Analytics for Mental States Assessment – A Review}, 
  year={2021},
  volume={},
  number={},
  pages={2266-2271},
  doi={10.1109/SMC52423.2021.9658674}}

@article{bend2026comparing,
  title={Comparing Eye-Tracking Metrics with the Driver Activity Load Index},
  author={Bend, Julia and G{\"o}dker, Markus and Banach, Elise Sophie and Franke, Thomas},
  journal={Journal of Eye Movement Research},
  volume={19},
  number={2},
  pages={28},
  year={2026},
  publisher={MDPI}
}

@article{yang2025,
author = {Yang, Tongyun and Regmi, Bishwas and Du, Lingyu and Bulling, Andreas and Zhang, Xucong and Lan, Guohao},
title = {Through the Eyes of Emotion: A Multi-faceted Eye Tracking Dataset for Emotion Recognition in Virtual Reality},
year = {2025},
issue_date = {September 2025},
publisher = {Association for Computing Machinery},
address = {New York, NY, USA},
volume = {9},
number = {3},
url = {https://doi.org/10.1145/3749545},
doi = {10.1145/3749545},
journal = {Proc. ACM Interact. Mob. Wearable Ubiquitous Technol.},
month = sep,
articleno = {143},
numpages = {41}
}

@inproceedings{behroozi2018can,
  title={Can we predict stressful technical interview settings through eye-tracking?},
  author={Behroozi, Mahnaz and Parnin, Chris},
  booktitle={Proceedings of the workshop on eye movements in programming},
  pages={1--5},
  year={2018}
}

@article{andersson2017one,
  title={One algorithm to rule them all? An evaluation and discussion of ten eye movement event-detection algorithms},
  author={Andersson, Richard and Larsson, Linnea and Holmqvist, Kenneth and Stridh, Martin and Nystr{\"o}m, Marcus},
  journal={Behavior research methods},
  volume={49},
  number={2},
  pages={616--637},
  year={2017},
  publisher={Springer}
}

@inproceedings{salvucci2000identifying,
  title={Identifying fixations and saccades in eye-tracking protocols},
  author={Salvucci, Dario D and Goldberg, Joseph H},
  booktitle={Proceedings of the 2000 symposium on Eye tracking research \& applications},
  pages={71--78},
  year={2000}
}

@article{ahmadi2024quantifying,
  title={Quantifying workload and stress in intensive care unit nurses: preliminary evaluation using continuous eye-tracking},
  author={Ahmadi, Nima and Sasangohar, Farzan and Yang, Jing and Yu, Denny and Danesh, Valerie and Klahn, Steven and Masud, Faisal},
  journal={Human factors},
  volume={66},
  number={3},
  pages={714--728},
  year={2024},
  publisher={Sage Publications Sage CA: Los Angeles, CA}
}

@article{GADSBY2021102932,
title = {An international comparison of the self-reported causes of cyclist stress using quasi-naturalistic cycling},
journal = {Journal of Transport Geography},
volume = {91},
pages = {102932},
year = {2021},
issn = {0966-6923},
doi = {https://doi.org/10.1016/j.jtrangeo.2020.102932},
url = {https://www.sciencedirect.com/science/article/pii/S0966692320310097},
author = {April Gadsby and Marjan Hagenzieker and Kari Watkins}
}

@article{RIVERAOLSSON2023107007,
title = {Are bicycle streets cyclist-friendly? Micro-environmental factors for improving perceived safety when cycling in mixed traffic},
journal = {Accident Analysis \& Prevention},
volume = {184},
pages = {107007},
year = {2023},
issn = {0001-4575},
doi = {https://doi.org/10.1016/j.aap.2023.107007},
url = {https://www.sciencedirect.com/science/article/pii/S0001457523000544},
author = {Sara {Rivera Olsson} and Erik Elldér}
}

@INPROCEEDINGS{Jyotsna,
  author={Jyotsna, C. and Amudha, J.},
  booktitle={2018 International Conference on Advances in Computing, Communications and Informatics (ICACCI)}, 
  title={Eye Gaze as an Indicator for Stress Level Analysis in Students}, 
  year={2018},
  volume={},
  number={},
  pages={1588-1593},
  doi={10.1109/ICACCI.2018.8554715}}

@ARTICLE{nemcova,
  author={Němcová, Andrea and Svozilová, Veronika and Bucsuházy, Kateřina and Smíšek, Radovan and Mézl, Martin and Hesko, Branislav and Belák, Michal and Bilík, Martin and Maxera, Pavel and Seitl, Martin and Dominik, Tomáš and Semela, Marek and Šucha, Matúš and Kolář, Radim},
  journal={IEEE Transactions on Intelligent Transportation Systems}, 
  title={Multimodal Features for Detection of Driver Stress and Fatigue: Review}, 
  year={2021},
  volume={22},
  number={6},
  pages={3214-3233},
  doi={10.1109/TITS.2020.2977762}}

@article{hooge2023robust,
  title={How robust are wearable eye trackers to slow and fast head and body movements?},
  author={Hooge, Ignace TC and Niehorster, Diederick C and Hessels, Roy S and Benjamins, Jeroen S and Nystr{\"o}m, Marcus},
  journal={Behavior Research Methods},
  volume={55},
  number={8},
  pages={4128--4142},
  year={2023},
  publisher={Springer}
}

@article{franchak2021adapting,
  title={Adapting the coordination of eyes and head to differences in task and environment during fully-mobile visual exploration},
  author={Franchak, John M and McGee, Brianna and Blanch, Gabrielle},
  journal={PLoS one},
  volume={16},
  number={8},
  pages={e0256463},
  year={2021},
  publisher={Public Library of Science San Francisco, CA USA}
}

@article{von2020gaze,
  title={Gaze behavior during urban cycling: Effects of subjective risk perception and vista space properties},
  author={von St{\"u}lpnagel, Rul},
  journal={Transportation research part F: traffic psychology and behaviour},
  volume={75},
  pages={222--238},
  year={2020},
  publisher={Elsevier}
}

@article{von2020crash,
  title={Crash risk and subjective risk perception during urban cycling: Evidence for congruent and incongruent sources},
  author={von St{\"u}lpnagel, Rul and Lucas, Jonas},
  journal={Accident Analysis \& Prevention},
  volume={142},
  pages={105584},
  year={2020},
  publisher={Elsevier}
}

@article{aldred2018cycling,
  title={Cycling injury risk in London: A case-control study exploring the impact of cycle volumes, motor vehicle volumes, and road characteristics including speed limits},
  author={Aldred, Rachel and Goodman, Anna and Gulliver, John and Woodcock, James},
  journal={Accident Analysis \& Prevention},
  volume={117},
  pages={75--84},
  year={2018},
  publisher={Elsevier}
}


\end{document}